\documentclass[twocolumn,nofootinbib,aps,amsmath,superscriptaddress,
preprintnumbers,floatfix]{revtex4}

\usepackage{epsfig}
\usepackage{graphicx}
\usepackage[hypertex]{hyperref}

\def\lqcd{\Lambda_{\rm QCD}}
\newcommand{\MSb}{\overline{\rm MS}}
\newcommand{\as}{\alpha_s}

\newcommand{\lamf}{\Lambda^{\!f}}

\newcommand{\ccm}{C_{\rm G}}
\newcommand{\bccm}{{\bar C}_{\rm G}}
\newcommand{\mcm}{\mu_{\rm G}^2}
\newcommand{\bRho}{\bar \Sigma_\rho}
\newcommand{\Rho}{\Sigma_\rho}

\newcommand{\nn}{\nonumber} 

\newcommand{\plus}{\ensuremath{\! + \!}}
\newcommand{\minus}{\ensuremath{\! - \!}}

\newcommand{\OMIT}[1]{}

\begin{document}

\preprint{\vbox{\hbox{MIT--CTP 4062}  \hbox{MPP-2009-156} 
\hbox{arXiv:0908.3189}
}}

\title{\boldmath $R$-evolution: Improving perturbative QCD \vspace{0.1cm}}

%\classification{11.10.Hi, 11.10.Jj, 11.15.Bt, 11.15.Me, 11.25.Db, 12.38.Aw, 12.38.Bx, 12.38.Cy, 12.39.Hg, 12.40.Yx, 14.40.Lb, 14.40.Nd, 14.65.Dw, 14.65.Fy, 14.65.Ha}

%\keywords{renormalization group, renormalons, chromomagnetic}

\author{Andr\'e H. Hoang}
  \affiliation{Max-Planck-Institut f\"ur Physik (Werner-Heisenberg-Institut), 
  F\"ohringer Ring 6, 80805 M\"unchen, Germany\vspace{0.0cm}}

\author{Ambar Jain}
\affiliation{Center for Theoretical Physics, Massachusetts Institute of
Technology, Cambridge, MA 02139\vspace{0.0cm}}

\author{Ignazio Scimemi}
\affiliation{Departamento de Fisica Teorica II,
Universidad Complutense de Madrid,
28040 Madrid, Spain\vspace{0.2cm}}

\author{Iain W.\ Stewart\vspace{0.2cm}}
\affiliation{Center for Theoretical Physics, Massachusetts Institute of
Technology, Cambridge, MA 02139\vspace{0.0cm}}

\begin{abstract}
  Perturbative QCD results in the $\MSb$ scheme can be dramatically improved by
  switching to a scheme that accounts for the dominant power law dependence on
  the factorization scale in the operator product expansion.  We introduce the
  ``MSR scheme'' which achieves this in a Lorentz and gauge invariant way.  The
  MSR scheme has a very simple relation to $\MSb$, and can be easily used to
  reanalyze $\MSb$ results.  Results in MSR depend on a cutoff parameter
  $R$, in addition to the $\mu$ of $\MSb$.  $R$ variations can be used to
  independently estimate i) the size of power corrections, and ii) higher order
  perturbative corrections (much like $\mu$ in $\MSb$).  We give two examples at
  three-loop order, the ratio of mass splittings in the $B^*$--$B$ and
  $D^*$--$D$ systems, and the Ellis-Jaffe sum rule as a function of momentum
  transfer $Q$ in deep inelastic scattering. Comparing to data, the perturbative
  MSR results work well even for $Q\sim 1\,{\rm GeV}$, and power corrections are
  reduced compared to $\MSb$.

\end{abstract}

\maketitle

\begin{center} {\bf Introduction and Formalism} \end{center} 
\vspace{-0.2cm} %\\[-10pt]

The operator product expansion (OPE) is an important tool for QCD. In hard
scattering processes two important scales are $Q$, a large moment transfer or mass,
and $\lqcd$, the scale of nonperturbative matrix elements.  The Wilsonian OPE
introduces a factorization scale $\lamf$, where $\lqcd < \lamf < Q$, and expands
in $\lqcd/Q$.  Consider a dimensionless observable $\sigma$ whose OPE is
\begin{align} \label{Wope}
 \sigma &= C_0^W(Q,\lamf ) \theta_0^W\!(\lamf)  
  + {C_1^W(Q,\lamf)}\, \frac{\theta_1^W\!(\lamf)}{Q^p}
 + \ldots .
% \sigma &= C_0^W(Q,\lamf ) \theta_0^W(\lamf,\lqcd)  \nn\\
% & + {C_1^W(Q,\lamf)}\, \theta_1^W(\lamf,\lqcd)/{Q^p}
% + \ldots .
\end{align}
The $C_{0,1}^W$ are dimensionless Wilson coefficients containing contributions
from momenta $k>\lamf$ with perturbative expansions in $\alpha_s$, and
$\theta_{0,1}^W=\langle {\cal O}_{0,1}\rangle_W$ are non-perturbative matrix
element with mass dimensions $0$ and $p$, containing contributions from
$k<\lamf$.  If $C_{0,1}^W(Q,\lamf)$ are expanded they contain an infinite series
of terms, $(\lamf/Q)^n$, modulo $\ln^m(\lamf/Q)$ terms, and this reflects the
fact that $C_{0,1}^W$ only include contributions from momenta $k>\lamf$.  The
Wilsonian OPE provides a clean separation of momentum scales, but can be
technically challenging to implement. In particular, it is difficult to define
$\lamf$ and retain gauge symmetry and Lorentz invariance, and perturbative
computations beyond one-loop are atrocious.

A popular alternative is the OPE with dimensional regularization and
the $\MSb$ scheme, which preserves the symmetries of QCD and provides
powerful techniques for multiloop computations. In this case
Eq.~(\ref{Wope}) becomes
\begin{align} \label{MSbope}
 \sigma &= \bar C_0(Q,\mu)\bar\theta_0(\mu)
  + {\bar C_1(Q,\mu)}
 \frac{\bar\theta_1(\mu)}{Q^p}
 + \ldots\,,
% \sigma &= \bar C_0(Q,\mu)\bar\theta_0(\mu,\Lambda_{\rm QCD})\: \nn\\
%  &+ {\bar C_1(Q,\mu)}\:
% \bar\theta_1(\mu,\Lambda_{\rm QCD})/{Q^p}
% + \ldots\,,
\end{align}
where $\mu$ is the renormalization scale and bars are used for $\MSb$
quantities.  In $\MSb$ the $\bar C_i$ are simple series in $\alpha_s$,
\begin{align} \label{Ci}
  \bar C_i(Q,\mu) & 
  =\! 1+ \sum_{n=1}^\infty b_n\big(\frac{\mu}{Q}\big) \:
   \Big[\frac{ \alpha_s(\mu)}{(4\pi)}\Big]^n \,,
\end{align}
with coefficients $b_n(\mu/Q)=\sum_{k=0} b_{nk} \ln^k(\mu/Q)$ containing only
$\ln\mu/ Q$. We will always rescale $\sigma$ and the matrix elements $\bar
\theta_i$ such that $\bar C_i=1$ at tree level.  In $\MSb$ all power law
dependence on $Q$ is manifest and unique in each term of Eq.~(\ref{MSbope}).
Also simple renormalization group equations in $\mu$, like $d \ln \bar
C_0(Q,\mu)/d\ln\mu = \bar\gamma[\alpha_s(\mu)]$, can be used to sum large logs
in Eq.~(\ref{MSbope}) if $Q\gg \Lambda_{\rm QCD}$.

$C_i^W(Q,\lamf)$ and $\bar C_j(Q,\mu)$ are related to each other in perturbation
theory, so Eqs.~(\ref{Wope}) and (\ref{MSbope}) are just the same OPE in two
different schemes. The renormalization scale $\mu$ in $\MSb$ plays the role of
$\lamf$.  This is precisely true for logarithmic contributions, $\ln\mu
\leftrightarrow \ln\lamf$, and here the Wilsonian picture of scale separation in
$\bar C_i$ and $\bar \theta_i$ carries over.  However, the same is not true for
power law dependences on $\lamf$.  $\MSb$ integrations are carried out over all
momenta, so the $\bar C_i$ actually contain some contributions from arbitrary
small momenta, and the $\bar \theta_i$ have contributions from arbitrary large
momenta.  For the power law terms there is no explicit scale separation in
$\MSb$, and correspondingly no powers of $\mu$ appear in Eq.~(\ref{Ci}).  While
this simplifies higher order computations, it is known to lead to factorial
growth in the perturbative coefficients. For $\bar C_0$, one has $b_{n+1}(\mu/Q)
\simeq (\mu/Q)^p\, n!\, [2\beta_0/p]^n Z$ at large $n$~\cite{Beneke:1998ui}, for
constant $Z$.  In practice this sometimes leads to poor convergence already at
one or two loop order in QCD.  This poor behavior is canceled by corresponding
instabilities in $\bar\theta_1$, and is referred to as an order-$p$ infrared
renormalon in $\bar C_0$ canceling against an ultraviolet renormalon in
$\bar\theta_1$~\cite{Mueller:1984vh,Luke:1994xd,Maiani:1991az}.  The
cancellation reflects the fact that the $\MSb$ OPE does not strictly separate
momentum scales.

The OPE can be converted to a scheme that removes this poor behavior, but still
retains the simple computational features of $\MSb$. Consider defining a new
``$R$-scheme'' for $C_0$ by subtracting a perturbative series
\begin{align} \label{C0R}
 C_0(Q,R,\mu) &= \bar C_0(Q,\mu) - \delta C_0(Q,R,\mu) \,,\nn\\
 \delta C_0(Q,R,\mu) &= \Big(\frac{R}{Q}\Big)^p \sum_{n=1}^\infty
  d_{n}\big(\frac{\mu}{R}\big) \:
  \Big[\frac{ \alpha_s(\mu)}{(4\pi)}\Big]^n \,,
%
%  C_0(Q,R,\mu) &= \bar C_0(Q,\mu) - \Big(\frac{R}{Q}\Big)^p \sum_{n=0}^\infty
%   a_{n}(\frac{\mu}{R}) \frac{ \alpha_s^n(\mu)}{(4\pi)^n}\\
%  &= \bar C_0(Q,\mu) - \Big(\frac{R}{Q}\Big)^p \sum_{n,k=0}^\infty
%   a_{nk} \frac{ \alpha_s^n(\mu)}{(4\pi)^n} \ln^k\!\frac{\mu}{R}  \nn.
\end{align}
with $d_n(\mu/R) = \sum_{k=0} d_{nk} \ln^k(\mu/R)$.  If for large $n$ the
coefficients $d_{n}$ are chosen to have the same behavior as $b_{n}$, so
$d_{n+1}(\mu/Q)\simeq (\mu/R)^p n!  [2\beta_0/p]^n Z$, then the factorial growth
in $\bar C_0(Q,\mu)$ and $\delta C_0(Q,R,\mu)$ cancel,
\begin{align} \label{noRenormalon}
 C_0(Q,R,\mu) \sim \Big[  \frac{\mu^p}{Q^p}\: 
   \!\!-\! \frac{R^p}{Q^p}\,\frac{\mu^p}{R^p} \Big] \sum_n
   n!  \Big[\frac{2\beta_0}{p}\Big]^n Z \,.
\end{align}
Thus the $R$-scheme introduces power law dependence on the cutoff, $(R/Q)^p$, in
$C_0(Q,R,\mu)$, which captures the dominant $(\lamf/Q)^p$ behavior of the
Wilsonian $C_0^W$. In practice this improves the convergence in $C_0$ even at
low orders in the $\alpha_s$ series.  The dominant effect of this change is
compensated by a scheme change to $\bar\theta_1$, $\bar\theta_1(\mu)=
\theta_1(R,\mu) - [Q^p \delta C_0(Q,R,\mu)] \bar \theta_0(\mu)$, and the new
$\theta_1$ will exhibit improved stability. In the $R$-scheme the OPE becomes
\begin{align} \label{Rope}
  \sigma &= C_0(Q,R,\mu) \bar\theta_0(\mu) 
    + \bar C_1(Q,\mu) {\theta_1(R,\mu)}/{Q^p}  \nn\\
   &\quad + \bar C_1^{\prime}(Q,\mu) {\theta_1'(R,\mu)}/{Q^p}
    +\ldots  ,
\end{align}
where $\theta_1'=[Q^p \delta C_0]\bar\theta_0$ and $\bar C_1'=1-\bar C_1\sim
\alpha_s$. Both $C_0$ and $\theta_1$ are free of order-$p$ renormalons. The
series in $\bar C_1'\, \theta_1'$ is Borel summable. In all examples below $\bar
\theta_0$ is also renormalon free.  The above procedure may be repeated for
higher renormalons and the higher power terms in the OPE indicated by ellipses,
to improve the behavior of these terms as well. At the order at which we work,
we will consistently set $\bar C_1=1$ and drop $\theta_1'$ in the following.

To setup an appropriate R-scheme it remains to define the $d_n$. In the
renormalon literature such scheme changes are well known for masses. For OPE
predictions a ``renormalon subtraction'' (RS) scheme has been implemented in
Ref.~\cite{Campanario:2005np}. In RS an approximate result for the residue of
the leading Borel renormalon pole is used to define the $d_n$, which adds a
source of uncertainty.

For our analysis we define the ``MSR'' scheme for $C_0$ by simply taking the
coefficients of the subtraction to be exactly the $\MSb$ coefficients. In
general it is more convenient to use $\ln \bar C_0$ rather than $\bar C_0$,
since this simplifies renormalization group equations.  Writing the series as
\begin{align} 
 \ln\bar C_0(Q,\mu) = \sum_{n=1}^\infty  a_n(\mu/Q) \Big[\frac{
   \alpha_s(\mu)}{(4\pi)}\Big]^n ,
\end{align}
with $a_n(\mu/Q) =\sum_{k=0} a_{nk}\ln^k\mu/Q$ we define the MSR scheme by the
series
\begin{align} \label{MSRln}
 \ln C_0(Q,R,\mu) \equiv \sum_{n=1}^\infty \Big\{
   a_n\big(\frac{\mu}{Q}\big) \!-\! 
   \frac{R^p}{Q^p} a_n\big(\frac{\mu}{R}\big) \Big\}
   \frac{ \alpha_s^n(\mu)}{(4\pi)^n} \,.
% \ln C_i(Q,R,\mu) \equiv \sum_{n=1}^\infty \Big\{
%   a_n\big(\frac{\mu}{Q}\big) \!-\! 
%  \Big(\frac{R}{Q}\Big)^p a_n\big(\frac{\mu}{R}\big) \Big\}
%   \Big[\frac{ \alpha_s(\mu)}{(4\pi)}\Big]^n .
\end{align}
This definition still cancels the order-$p$ renormalon for large $n$, as in
Eq.~(\ref{noRenormalon}).  It yields the very simple relation
\begin{align} \label{MSR}
  C_0(Q,R,\mu) = \bar C_0(Q,\mu) \big[ \bar
  C_0(R,\mu)\big]^{-(R/Q)^p} \,,
\end{align}
which must be expanded order-by-order in $\alpha_s(\mu)$ to remove the
renormalon.  Thus the coefficient $C_0(Q,R,\mu)$ for the MSR scheme is obtained
directly from the $\MSb$ result. Note $C_0(Q,Q,\mu) = 1$ to all orders. The
appropriate $p$ is obtained from the $\MSb$ OPE by
$p$=dimension($\bar\theta_1)-$ dimension($\bar\theta_0$).  MSR preserves gauge
invariance, Lorentz symmetry, and the simplicity of $\MSb$.

The appropriate values for $R$ in Eqs.~(\ref{C0R},\ref{Rope},\ref{MSR}) are
constrained by power counting and the structure of large logs in the OPE.  The
power counting $\bar\theta_1 \sim \Lambda_{\rm QCD}^p$ implies $\theta_1\sim
\Lambda_{\rm QCD}^p$, so for the matrix element we need $R\sim\mu\gtrsim
\Lambda_{\rm QCD}$ (meaning a larger value where perturbation theory for the OPE
still converges), which minimizes $\ln(\mu/\Lambda_{\rm QCD})$ and $\ln(\mu/R)$
terms in $\theta_1(R,\mu,\lqcd)$.  On the other hand, $C_0(Q,R,\mu)$ has
$\ln(\mu/Q)$ and $\ln(\mu/R)$ terms, and for $R\sim \Lambda_{\rm QCD}$ no choice
of $\mu$ avoids large logs.  For $R\sim\mu\sim Q$ we can minimize the logs in
$C_0(Q,R,\mu)$, but not in $\theta_1(R,\mu,\lqcd)$.  When the OPE is carried out
in $\overline{\rm MS}$ this problem is dealt with using a $\mu$-RGE to sum large
logs between $Q$ and $\Lambda_{\rm QCD}$.  For MSR we must use $R$-evolution, an
RGE in the $R$ variable~\cite{Hoang:2008yj}.  The appropriate R-RGE is
formulated with $\mu=R$ to ensure there are no logs in the anomalous dimension.
For $C_0$,
\begin{align}\label{RRGE}
R\frac{d}{dR} \ln C_0(Q,R,R) =  \bar \gamma[\alpha_s(R)] 
  - \Big(\frac{R}{Q}\Big)^p \gamma[\alpha_s(R)],
\end{align}
where $\bar \gamma[\alpha_s]=\sum_{n=0}^\infty \bar\gamma_n
[\alpha_s(R)/4\pi]^{n+1}$ and $\gamma[\alpha_s]=\sum_{n=0}^\infty \gamma_n
[\alpha_s(R)/4\pi]^{n+1}$ are the $\MSb$ and $R$ anomalous dimensions.  Here
$\gamma_{n-1} = p a_{n0} - 2 \sum_{m=1}^{n-1} m \, a_{m0} \, \beta_{n-m-1}$ and
we are using the $\MSb$ $\beta$-function $\mu d/d\mu \alpha_s(\mu) =
-\alpha_s^2(\mu)/(2\pi) \sum_{n=0}^\infty \beta_n [\alpha_s(\mu)/4\pi]^n$. The
choice in Eq.~(\ref{MSRln}) keeps Eq.~(\ref{RRGE}) simple. In cases where
$\bar\gamma$ is absent we expect Eq.~(\ref{RRGE}) to converge to lower scales
due to the $(R/Q)^p$ factor multiplying $\gamma$. For $R_1>R_0$ the solution of
Eq.~(\ref{RRGE}) is [$U_\mu=U_\mu(R_1,R_0)$]
\begin{align}   \label{Csoln}
  & C_0(Q,R_0,R_0) = C_0(Q,R_1,R_1) \, U_R(Q,R_1,R_0)  \, U_\mu \,, 
\end{align}
where $U_\mu$ is a usual $\MSb$ evolution factor and $U_R$ is the $R$-evolution.
For $p=1$ the complete solution for $U_R$ was obtained in
Ref.~\cite{Hoang:2008yj}. It is straightforward to generalize this to any $p$.
At N${}^{k+1}$LL order the (real) result is
\begin{align}  \label{UR}
 &U_R(Q,R_1, R_0) = \exp \bigg\{ \Big(\frac{\Lambda^{(k)}_{\rm QCD}}{Q}\Big)^p \sum_{j=0}^k  S_j\, (-p)^j 
  e^{i\pi p \hat b_1}\nn \\
  &\times  p^{(p  \hat b_1)}   \big[ \Gamma(-p \hat b_1\minus j,p t_0) 
  - \Gamma(-p \hat b_1 \minus j, p t_1) \big] \bigg \},
\end{align}
with $\Gamma(c,t)$ the incomplete gamma function and $t_{0,1} =
-2\pi/(\beta_0\as(R_{0,1}))$.  $\lqcd^{(0)} = R e^{t}$, $\lqcd^{(1)} = R
e^{t}(-t)^{\hat b_1}$, and $\lqcd^{(2)} = R e^{t}(-t)^{\hat b_1} e^{-\hat
  b_2/t}$ are evaluated at a large reference $R$ with $t=-2\pi/(\beta_0\as(R))$,
and $\hat b_1 = \beta_1/(2\beta_0^2)$, $\hat
b_2=(\beta_1^2-\beta_0\beta_2)/(4\beta_0^4)$, and $\hat
b_3=(\beta_1^3-2\beta_0\beta_1\beta_2+\beta_0^2\beta_3)/(8\beta_0^6)$.  Defining
$\tilde \gamma_n = \gamma_n/(2\beta_0)^{n+1}$ the coefficients of $U_R$ needed
for the first three orders of $R$-evolution are
\begin{align} \label{Si}
&  S_0 = \tilde \gamma_0 \,,  
 \qquad 
   S_1 = \tilde \gamma_1 - (\hat b_1\plus p \hat b_2) \tilde \gamma_0\,,
   \\
& S_2 = \tilde \gamma_2 - (\hat b_1\plus p \hat b_2) \tilde \gamma_1
    +\big[ (1\plus \hat p b_1)\hat b_2 +p(p\hat b_2^2 \plus \hat b_3)/2 \big]
  \tilde \gamma_0 \, .
  \nn
\end{align}
Eq.~(\ref{Rope}) becomes $C_0(Q,R_1,R_1) U_R(Q,R_1,R_0)  U_\mu(R_1,R_0)$ $\times
\theta_0(R_0)+ \theta_1(R_0,R_0)/Q^p$, and this result %which 
sums logs between
$R_1\sim Q$ and $R_0\sim \lqcd$. This gives natural $R$ scales for 
coefficients and matrix elements in the OPE.

%\vspace{0.2cm}
%\noindent {\bf Heavy Meson Mass Splittings in MSR}\\[-8pt]
\begin{center} {\bf Heavy Meson Mass Splittings in MSR} \end{center} 
\vspace{-0.2cm}

The $\MSb$ OPE for the mass-splitting of heavy mesons, $\Delta m_H^2 =
m_{H^\ast}^2-m_H^2 $ for $H=B,D$, is $\Delta m_H^2 = \bccm(m_Q,\mu)\, \mcm(\mu)
+ \sum_i \bar C_i(m_Q,\mu)\, 2\rho_i^3(\mu)/(3m_Q) + {\cal O}(\lqcd^3/m_Q^2)$,
where $m_Q=m_{b}$ or $m_c$. Here $\mcm = -\langle B_v| \bar h_v g
\sigma_{\mu\nu}G^{\mu\nu}h_v|B_v\rangle/3$ is the matrix element of the
chromomagnetic operator, and $\rho_i^3$ for $i=\pi
G,A,LS,\bar\Lambda G$ are ${\cal O}(\lqcd^3)$ matrix
elements~\cite{Grozin:1997ih}, with $\rho^3_{\bar\Lambda G}(\mu) =
(3/2)\bar\Lambda\mcm(\mu)$. At the order of our analysis tree level values for
the $\bar C_i$ suffice, so with $\bRho(\mu) = (2/3) \big [\rho_{\pi
  G}^3(\mu)+\rho_{A}^3(\mu)-\rho_{LS}^3(\mu) + \rho_{\bar \Lambda G}^3(\mu) \big
]$,
\begin{align}  \label{DelmHope}
 \Delta m_H^2 
 &= \bccm(m_Q,\mu)\, \mcm(\mu) + 
 {\bar\Sigma_\rho(\mu)}/{m_Q} + \ldots \,.
\end{align} 
Taking the ratio of mass splittings $r=\Delta m_B^2/\Delta m_D^2$ gives
\begin{align} \label{rMSb}
 r & %= \frac{\Delta m_B^2}{\Delta m_D^2}
%= \frac{m_{B^*}^2-m_B^2}{m_{D^*}^2-m_D^2} 
%  =  \frac{\bar C_G(m_b,\mu) \mu_G^2(\mu) 
%  + \bRho(\mu)/m_b}{\bar C_G(m_c,\mu) \mu_G^2(\mu) 
%  + \bRho(\mu)/m_c} + \ldots \nn \\
= \frac{\bar C_G(m_b,\mu)}{\bar C_G(m_c,\mu)} 
+ \frac{\bRho(\mu)}{\mu_G^2(\mu)} 
 \Big(\frac{1}{m_b}-\frac{1}{m_c}\Big) + \ldots.
\end{align}
The first term in this OPE gives a purely perturbative prediction for $r$.
$\bccm$ is known to suffer from an ${\cal O}(\lqcd/m_Q)$ infrared renormalon
ambiguity~\cite{Grozin:1997ih}, with a corresponding ambiguity in $\bRho(\mu)$.
The three-loop computation of Ref.~\cite{Grozin:2007fh} yields, $r = 1
-0.1113|_{\alpha_s} -0.0780|_{\alpha_s^2} -0.0755|_{\alpha_s^3}$ at fixed order
with $\mu=m_c$, and $r = (0.8517)_{\rm LL} + (-0.0696)_{\rm \Delta NLL} +
(-0.0908)_{\rm \Delta NNLL}$ in RGE-improved perturbation theory, with no sign
of convergence in either case.  In $\MSb$ these leading predictions are unstable
due to the $p=1$ renormalon in $\bar C_G$.

Lets examine the analogous result in the MSR scheme
\begin{align}\label{eqn:OPE}
  \Delta m_H^2
  & = \ccm(m_Q,R,\mu) \mcm(\mu) + \frac{\Rho(R,\mu)}{m_Q} + \ldots \,.
\end{align}
Since $p=1$ the MSR definition in Eq.~(\ref{MSR}) gives
\begin{equation} \label{eqn:CRF}
\ccm(m_Q,R,\mu)\equiv \bccm(m_Q,\mu) [\bccm(R,\mu)]^{-R/m_Q} \,,
\end{equation}
where $\bccm(m,\mu)$ is obtained from Ref.~\cite{Grozin:2007fh} and we expand in
$\alpha_s(\mu)$.  The OPE in MSR at a scale $R_0\gtrsim\lqcd$ gives
\begin{align}
  & r = \frac{\ccm(m_b,R_0,R_0)}{\ccm(m_c,R_0,R_0)} +
  \frac{\Rho(R_0,R_0)}{\mu_G^2(R_0)}
  \Big(\! \frac{1}{m_b} \minus \frac{1}{m_c} \! \Big) \,.
\end{align}
Large logs in $C_G(m_Q,R_0,R_0)$ can be summed with the R-RGE in
Eqs.~(\ref{Csoln}--\ref{Si}). For simplicity we integrate out the $b$ and
$c$-quarks simultaneously at a scale $R_1 \simeq \sqrt{m_b m_c}\gg R_0\simeq
\lqcd$. This scale for $R_1$ keeps $\ln(R_1/m_{b,c})$ small.  With
R-evolution and $U_R$ from Eq.~(\ref{UR}) we have
\begin{align} \label{rR}
 r&= \frac{\ccm(m_b,R_1,R_1\!)U_R(m_b,R_1,R_0)}
  {\ccm(m_c,R_1,R_1\!)U_R(m_c,R_1,R_0)}
   \\
   &\ \ \ 
+ \frac{\Rho(R_0,R_0)}{\mu_G^2(R_0)}
%   \frac{(m_c\!-\!m_b)}{m_bm_c}
   \Big(\! \frac{1}{m_b} \minus \frac{1}{m_c} \! \Big )
  . \nn 
\end{align} 
This expression is independent of $R_1$ and $R_0$.  Order-by-order, varying
$R_1$ about $\sqrt{m_bm_c}$ yields an estimate of higher order perturbative
uncertainties, much like varying $\mu$ in $\MSb$.  For $R_0$ the dependence
cancels between the first term in $r$ and the $\Sigma_\rho$ power
correction. In MSR the term $\Sigma_\rho(R_0,R_0)$ is $\sim \lqcd^3$ and can be
positive or negative. One may expect that there is a value of $R_0$ where
$\Sigma_\rho(R_0,R_0)$ vanishes.  Thus keeping only the first term in
Eq.~(\ref{rR}) and varying $R_0\gtrsim \lqcd$ yields an estimate for the size of
this power correction.  This technique goes beyond the dimensional analysis
estimates used in $\MSb$.

\begin{figure}[t!]
\centering
\includegraphics[width=1.\columnwidth]{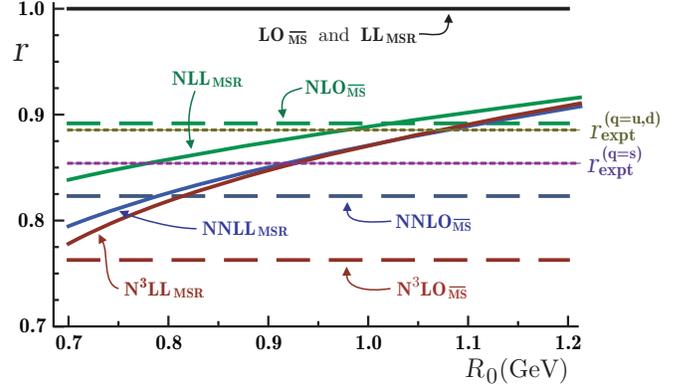}
%       \label{fig:chromo-RF-new-2}
%\end{figure}
%\begin{figure}[t]
%  \centerline{\includegraphics[width=1 \columnwidth]{figs/chromo-RF-new-2}}
\vspace{-0.6cm} 
\caption{\label{fig:chromo-RF} Perturbative predictions at leading order in
  $1/m_Q$ for the ratio $r$ of the $B$-$B^\ast$ and $D$-$D^\ast$ mass splittings
  in the MSR-scheme (solid) versus $\MSb$ (dashed).  The $R_0$ dependence of the
  solid red curve provides an estimate for the power correction, independent of
  the comparison with the experimental data.  Neither $R_1$ nor $\mu$ variation
  is shown in the figure.}  \vspace{-0.6cm}
\end{figure}
Fig.~\ref{fig:chromo-RF} gives perturbative predictions for $r$ at different
orders using the first terms in Eqs.~(\ref{rMSb},\ref{rR}) with $m_b = 4.7$ GeV,
$m_c = 1.6$ GeV, $\as(\sqrt{m_bm_c}) = 0.2627$ and the 4-loop $\beta$-function.
The solid lines are from the MSR scheme, plotted as functions of $R_0$.  The
dashed lines are the fixed order $\MSb$ results with $\mu=\sqrt{m_b m_c}$. The
MSR results exhibit a dramatic improvement in convergence over $\MSb$ for a wide
range of $R_0$ values.  Varying $R_1=\sqrt{m_bm_c}/2$ to $2\sqrt{m_bm_c}$ at
N${}^3$LL(MSR) gives $\Delta r\simeq \pm 0.008$, which is a significant
improvement over $\mu$ variation in the same range for N${}^3{\rm LO}$${(\MSb)}$
where $\Delta r\simeq \pm 0.068$.  The MSR results converge to an $R_0$
dependent curve, whose dependence cancels against $\Rho(R_0,R_0)$, so the
residual $R_0$ dependence provides a method to estimate the size of this power
correction. The range $R_0 = 0.7\,{\rm GeV}$ to $1.2\,{\rm GeV}$ keeps $R_0$
below $m_c$ and above $\lqcd$ and yields
\begin{equation}
r = 0.860 \pm (0.065)_{\Sigma_\rho} \pm (0.008)_{\rm pert.} \,.
\end{equation}
This estimate for the $\Sigma_\rho$ power correction in MSR is in good agreement
with experiment, $r_{\rm expt}=0.886$ ($D_{u,d}^{(*)}$, $B_{u,d}^{(*)}$) and
0.854 ($D_s^{(*)}$, $B_s^{(*)}$).  MSR achieves a convergent perturbative
prediction for $r$ at leading order in the OPE, and a $1/m_Q$ power correction
of moderate size, $\sim 0.065$, significantly smaller than the dimensional
analysis estimate of $\Lambda_{\rm QCD}(1/m_{c}-1/m_b)\sim 0.2$ in $\MSb$.

%%%%%%%%%%%%%%%%%%%%%%%%%%%%%%%%%%%%%%%%%%%%%
%%%%%%%%%%%%%%%%%%%%%%%%%%%%%%%%%%%%%%%%%%%%%
\begin{center} {\bf Ellis-Jaffe sum rule in MSR} \end{center} 
\vspace{-0.2cm}

\begin{figure}[t!]
  \centerline{\includegraphics[width=1 \columnwidth]{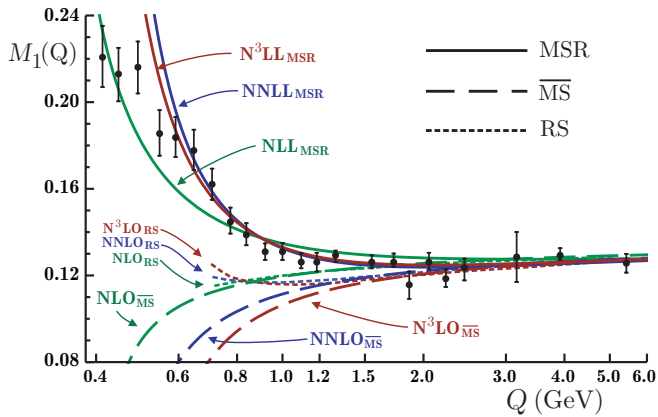}}
 \vspace{-0.2cm}
 \caption{\label{fig:EJSR2} Perturbative results for the Ellis-Jaffe sum rule in
   the MSR, RS, and $\MSb$ schemes, at leading order in $1/Q$. For all curves
   the one parameter, $\hat a_0$, is fixed by data at $Q\simeq 5\,{\rm GeV}$.
 }
 \vspace{-0.2cm}
\end{figure}

\begin{figure}[t!]
  \centerline{\includegraphics[width=1 \columnwidth]{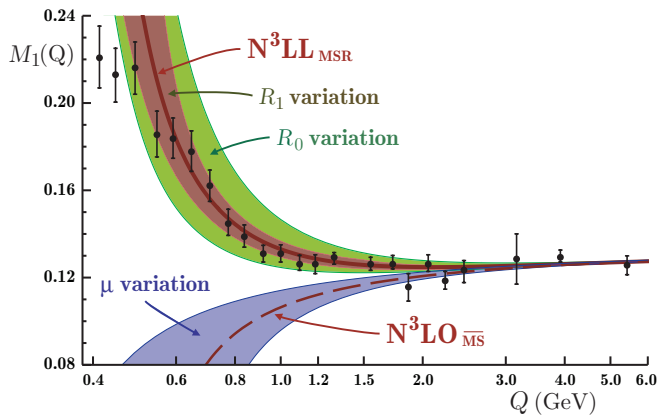}}
 \vspace{-0.2cm}
 \caption{\label{fig:EJSR} Uncertainty estimates in the MSR scheme and $\MSb$
   scheme for the Ellis-Jaffe sum rule at leading order in $1/Q$.  }
 \vspace{-0.2cm}
\end{figure}
%\noindent

In $\MSb$ the Ellis-Jaffe sum rule~\cite{Ellis:1973kp} for the proton in DIS
with momentum transfer $Q$ is $M_1(Q)=\big[\bar C_B(Q,\mu)\,\theta_B+ \bar
C_{0}(Q,\mu) {\hat a_0}/{9}\big] + {\bar \theta_1(\mu)}/{Q^2} $. $\bar C_{B,0}$
are known at 3 loops~\cite{Larin:1997qq}.  The two leading order terms are
written so that both coefficients and matrix elements are separately
$\mu$-independent: $\theta_B=g_A/12+a_8/36$ is given by the axial couplings
$g_A=1.2694$ and $a_8=0.572$ for the nucleon and hyperon, while $\hat a_0$ is a
$Q$ independent $\MSb$ matrix element.  $\bar\theta_1$ denotes all $1/Q^2$ power
corrections with their Wilson coefficients at tree level.  The $\MSb$
coefficients are affected by a $p=2$ renormalon~\cite{Broadhurst:1993ru}, which
is removed in the MSR scheme.  Eq.~(\ref{MSR}) gives [$i=B,0$]
\begin{align} \label{eq:ejf}
C_i(Q,R,R)\equiv \bar C_i(Q,R) [\bar C_i(R,R)]^{-R^2/Q^2} \,.
\end{align}
\phantom{text}

With $R$-evolution the MSR OPE prediction is
\begin{align}
& M_1(Q) =
 \big[C_B(Q,R_1,R_1) U^{B}_R(Q,R_1,R_0)\theta_B \\
& +C_0(Q,R_1,R_1) U^{0}_R(Q,R_1,R_0) \hat a_0/9 \big] +
\theta_1(R_0,R_0)/Q^2 \nn ,
\end{align}
where $U_R^{B,0}$ are given by Eq.~(\ref{UR}) with $p=2$ and the corresponding
$(a_{n0})^{B,0}$ determine the appropriate $(\gamma_n)^{B,0}$.

Figures~\ref{fig:EJSR2},\ref{fig:EJSR} show perturbative predictions for the
Ellis-Jaffe sum rule at leading power in $1/Q$, compared with proton data from
Ref.~\cite{Osipenko:2005nx}. We use $\alpha_s(4\,{\rm GeV})=0.2282$, and the
4-loop $\beta$ with $4$ flavors. In Fig.~\ref{fig:EJSR2}, we show order-by-order
results for the $\MSb$ scheme at $\mu=Q$, and for the resummed MSR scheme with
$R_1=Q$ and $R_0=0.9$ GeV.  We fix $\hat a_0=0.141$ so that $\MSb$ and MSR agree
with the data for $Q\simeq 5\,{\rm GeV}$.  $\MSb$ agrees well with the data for
large $Q$, but turns away at $Q \lesssim 2$ GeV and no longer converges. In
contrast the MSR results still converge quickly and exhibit excellent agreement
with the data over a wide range of $Q$ values.  The NLL MSR result already has
the right curvature, and the NNLL and N$^{3}$LL curves further improve the
agreement.  We also include predictions in the RS scheme with subtraction scale
$\nu_f = 1.0\,{\rm GeV}$ from Fig.3d of Ref.~\cite{Campanario:2005np}, which
improve slightly over the $\MSb$ results, but may not be capturing the dominant
power law dependence on the factorization scale.  In Fig.~\ref{fig:EJSR} we show
uncertainties for three loop results in the $\MSb$ and MSR schemes.  The dashed
red curve is the $\MSb$ prediction, and the blue band estimates the higher-order
perturbative uncertainties varying $\mu$ in the range $\mu^{min}(Q)<\mu<2Q$. For
$Q>1.5\,{\rm GeV}$, $\mu^{min}=Q/2$, while for $Q<1.5$ GeV, $\mu^{min}= 1.3 Q/(1.1+
Q/(1\,{\rm GeV}))$.  The red solid line is the MSR prediction, the red band is the
perturbative uncertainty from varying $R_1$ in the same range as was done for
$\mu$ in $\MSb$, and the green band estimates the $1/Q^2$ power correction by
varying $R_0=0.7$ to $1.2 ~{\rm GeV}$.  Fig.~\ref{fig:EJSR} implies $-0.01\,{\rm
  GeV}^2\lesssim \theta_1(R_0,R_0)\lesssim 0.01\,{\rm GeV}^2$ in MSR, which is a
much smaller power correction than the $\sim 0.1$ GeV$^2$ estimate obtained from
naive dimensional analysis in $\MSb$.

%%%%%%%%%%%%%%%%%%%%%%%%%%%%%%%%%%%%%%%%%%%%%
%%%%%%%%%%%%%%%%%%%%%%%%%%%%%%%%%%%%%%%%%%%%%

%{\bf Acknowledgments:} 
This work was supported by the EU network, MRTN-CT-2006-035482 (Flavianet),
Spanish Ministry of Education, FPA2008-00592, the Office of Nuclear Physics of
the U.S.\ Department of Energy, DE-FG02-94ER40818, the Alexander von Humboldt
foundation, and the Max-Planck-Institut f\"ur Physik guest progam.

%%%%%%%%%%%%%%%%%%%%%%%%%%%%%%%%%%%%%%%%%%%%%
%%%%%%%%%%%%%%%%%%%%%%%%%%%%%%%%%%%%%%%%%%%%%

%\bibliographystyle{iain} 
%\bibliography{hjss2}

\end{document}